# Understanding Working from Home Practical Changes and Adaptations During the COVID-19 Pandemic

Jie Cai, Sarah J Ryu, Hyejin Hannah Kum-Biocca, Donghee Yvette Wohn
New Jersey Institute of Technology
New Jersey, United States
jie.cai@njit.edu; sr984@njit.edu; hyejin.kum-biocca@njit.edu; wohn@njit.edu

**Abstract**

*While much work focuses on the impacts of the pandemic on people's psychological and physical health, it is still unclear about the practical changes and adaptations. In this work, we interviewed 46 participants who were forced to work from home. Results show that there is an increased reliance on asynchronous communication, which slowed communication efficiency and decreased initiative to communicate. The home environment causes distraction from households and lacked facilities but is embraced by a group of people. Many people had to passively adapt to the communication and environmental changes and accept the limitations of technology, a situation that is not sustainable in the long run. We pointed out how technology can potentially play a larger role in supporting communication and coping with environmental changes in the future.*

## 1. Introduction

In March 2020, the governments mandated social distancing policies and locked down some states because of the COVID-19 pandemic; accordingly, many organizations had to request their employees to work from home to reduce the infection risk [1]. The world-spread pandemic with the government-imposed restriction has significantly changed the way people live, work, communicate, and learn, and these changes have become the new normal of daily life [2]. Suddenly, all employees become telecommuters with little or no preparation.

The enforced work from home transition increases our dependency on information systems and technology [3]. The lockdown accelerates the trend to migrate to telecommuting and appropriate technology use. One indicator of this was a sudden surge in video conferencing. The downloading of video conferencing software jumped 45% immediately after the early lockdown and reached a 90% average increase in downloads compared to pre-COVID demand [4], such as Zoom, Google, and Microsoft telework tools [5].

The lockdown due to pandemic offers a unique context that is significantly different from the previous remote work literature. The difference is that the work from home is enforced and applied to all, restricts mobility, and maintains social distancing among people; thus, the enforced work from home causes challenges on communication, environment negotiation, and well-being [3]. Telecommuters need to adapt quickly to non-conducive working spaces and unfamiliar digital technologies to maintain regular business. While much work focuses on the psychological impact on the forced telecommuters (e.g., see meta-analysis by [6]), there is less work investigating the practical transition and adaptation. Understanding these issues can help organizations and designers to design better intervention or solutions to overcome the challenges people faced during the process.

In this work, we interviewed 46 telecommuters working in the pandemic to understand the changes and their adaptation to these changes during the work practices. We found that the existing technology for communication slows the communication efficacy and reduces the initiative to have communication. The home environment is usually distracting, lacking facilities and equipment, but favored by those who have a workstation setup and enjoy the quiet and work-alone environment. The adaptations to the communication and environment changes include psychological (shift in mental state), physical (force to have more physical activities), and practical activities (improve personal skills and expand technology portfolio) with little technological intervention (planning technology to push self with a hard deadline). The adaptations can only overcome some of the communication changes, but not the environmental changes, requiring future investigation.

## 2. Related Work

### 2.1. Rapid Migration to Telecommuting

Since the arrival of networked computers, there has been a slow but upward rise in telecommuting and working from home. There was more than a 150% increase in telecommuting in the last 15 years [7]. More than 50% of companies allowed for remote work with a mix of about 30% of the employees who telecommute doing it full time, and the remainder was occasionally working from home. From 2010 to 2020, corporations reported an increase in over 400% in the number of employees working remotely at least once a week [8]. Most studies of telecommuting involved mixtures of full-time and part-time telecommuters. Under normal telecommuting, 23% of remote workers say they are willing to work longer hours at home than in the office [9]. Nevertheless, there are reports that the boundary between work and home time can be a problem for remote workers [10]. Telecommuters pre-COVID report concerns about the deterioration of relationship quality with co-workers, supervisors, and family [11].

The government lockdown and public health requirements for social distancing during COVID led to explosive growth in the existing movement towards telecommuting. The pandemic forced workers, for the most part, to all be full-time telecommuters. The trend before and during COVID isolation suggests that telecommuting may become a norm for many going forward. Some industry analysts predict an increase of 77% from 2019 to 2022. A survey during COVID reported that 55% of US workers want a mixture of home and office working [12]. A study of UK employers found that employers expect the proportion of regular home workers to double from 18% pre-pandemic to 37% post-pandemic [13].

The migration provides opportunities to technological development and raises new challenges. Remote work and work from home have drawn significant research attention in decades [14, 15, 16]. However, before the pandemic, work from home is organizational-specific and optional. The lockdown and social distancing policies force to almost all employed or self-employed workers to work from home without any discrimination [17]. During the transition, telecommuters are forced to use technologies in new ways to complete their tasks, interact with co-workers and colleagues, and arrange the home environment for working [3]. It is still, however, not clear what the practical changes are. Thus, we ask:

- RQ1: What are the changes in the forced telecommuters' work practices?

### 2.2. Impacts of Telecommuting Under COVID-19

The movement to remote, mediated interaction in telecommuting changes the nature of work practices and behaviors and attitudes toward work. The rapid movement to telecommuting appeared to accelerate these changes. A recent survey of 1,046 employers by the CIPD found 28% reporting that they increase in homeworking during lockdown had increased productivity or efficiency [13]. Notably, this number is less than pre-COVID data. COVID-induced changes suddenly disrupted workflows in organizations, leading to productivity impediments for some workflows not already structured for telecommuting. Under COVID-19, employers reported that the proportion of staff who worked from home all the time rose to 22% post-pandemic, compared with 9% previously. The speed of the transition was a vital issue [18]. A related issue to the sudden change was the need for some workers to suddenly learn new remote technologies and workflows [19]. The movement also causes psychological impacts. The forced and sudden migration to online interaction under COVID may have made some of these issues more pronounced. The movement towards telecommuting under COVID was accompanied by a great deal of uncertainty regarding personal health, job stability, and macro-economic outcomes [20], increased anxiety, stress, and depression [21, 6].

Much work explores the impact of the pandemic in various perspectives, such as occupational status [22, 23], economic uncertainty [20] work from home boundary [24], and the psychological issues related to physical and mental health [25, 26, 1, 6]. These studies also propose possible psychological and technological interventions to mitigate the negative impacts aforementioned [27, 6] such as how to adapt technology to overcome technological and contextual shortcomings [17]. In line with the proposition, we intend to explore the adaptation strategies and how they work to cope with the changes. Thus, we ask:

- RQ2: How do the telecommuters adapt work practices?

## 3. Method

### 3.1. Participant Recruitment

The projects and the interview protocol were reviewed and approved by the Institutional Review Board (IRB). We recruited adults in the U.S. who have to work from home due to the pandemic. We used a convenience sample - 18 were family, friends, or loose acquaintances of four research assistants and 12 students taking a computing-related course. We interviewed 46 participants total. A majority of participants were in their 20's (60.9%) ( $M$= 27.7, $range$= 19 to 55). There were 19 female (41.3%) and 22 male participants (47.8%). Most of them are Asian (50.0%), followed by White (21.7%), African American (10.9%), and Hispanic (6.5%). Five participants did not share their demographics (10.9%), labeled as N/A. Demographic details are summarized in the the Appendix.

### 3.2. Interview Process

We conducted most semi-structured interviews through remote calls via Discord, Skype, or Face-Time, lasting about 15 to 45 minutes. During the interview, we first asked general warm-up questions about their working routines and working environment such as "What does your current living situation look like?" and "Can you walk me through what you did yesterday?". Then, we asked more specific questions related to our research questions, along with follow-up questions about their work from home job experience and work collaboration experiences like "Can you think of someone whom you collaborated with remotely for work?" and follow-up questions about their personal thoughts on the transition and changed social situation like "How do you communicate with your co-workers?" and "How has the current situation affected your job performance?" At the end of the interview, we asked basic demographic questions about age, race, and gender and encouraged participants to share anything related to working from home that we did not ask during the interview.

### 3.3. Interview Analysis

All the interviews were audio-recorded, transcribed by speech recognition software, and then rectified by research assistants. We applied a thematic analysis to identify, analyze, organize, describe, and report themes [28] with an inductive coding process. First, we familiarized ourselves with all transcripts and organized the content on a spreadsheet by putting participants' relevant answers under our research questions. Second, each researcher openly coded several (three to six) transcripts each week for the first research question, then presented their codes at the weekly calibration meetings. During the meeting, we went through the quotes and related codes together. Each researcher presented and explained their codes for the rest of the team to review and ask any questions, until all the codes were consistent and finalized. For the first research question, we iteratively coded and held review meetings for about four weeks. We applied a similar process to code the second research question, which took about three weeks.

After we completed the coding, we imported all the codes into a collaborative whiteboard called Miro. On the whiteboard, each code was presented on a sticky note with a tag that indicated the participant number. All the researchers organized the notes simultaneously to interactively group codes together while having the voice call to discuss the themes. After we defined the themes, we went more in-depth and reviewed the codes in each theme to find any sub-themes. This process was completed within several weekly meetings and reflections. The last step was to look for representative quotes based on the codes and report in the result section.

## 4. Results

### 4.1. Practical Changes Due to Pandemic

The first research question investigates the changes people experienced while transitioning to working from home due to the COVID-19 pandemic. We delve into three categories related to changes in communication and two categories related to changes in the work environment.

#### 4.1.1. Adjustments in Overall Communication

**Rely on Asynchronous Communication** People who switched to working from home were using more asynchronous (written) communication methods (e.g., texting, messaging, emailing) alongside synchronous communication methods (e.g., voice calls, video calls). While they preferred physical, in-person interaction, the poor infrastructure prevented them from having synchronous live meetings all the time. Some needed to grow accustomed to relying more on asynchronous communication. P24 (26, M), a senior research analyst at a healthcare market research company, said, *"It definitely requires more written communication with my co-workers since we don't have the opportunity of being in the same place versus being able to walk over to someone at and conversing with them so that's a big hurdle especially with my team since we weren't really into the written communication before."*

**Lack of Initiative in Communication** People working from home faced a decrease in communication with coworkers as a result of seeing coworkers less.

P40 (N.A, N.A), a product owner in the customer technology team at a large energy company, felt that being at home provided less initiative to socialize with their coworkers: *"It's because you're home now all the time. And the only downside is I don't see my co-workers that much. So it's like we don't really talk every day."* There is less motivation to communicate online with coworkers outside of work, whereas in-person communication would be able to provide more flexibility for socialization (P15, 24, F).

The archived nature of asynchronous communication also reduced the initiative to interact with others. P45 (21, F), a student, found that being able to review audio-recorded lecture material multiple times resulted in asking less questions to her professor: *"In class you only get to hear his lecture once and so you can ask any questions, maybe you had a hard time hearing him. But now it would be like, 'Oh, I have that access to watch it over and we re-watch it again. Why would I ask him the question?' So there's less questions."*

**Slow Communication and Response** Being in the office provided physical cues that suggested a person's availability, which made it easier for people to start spontaneously interacting and collaborating with coworkers without needing to schedule a meeting. Now, people who work from home needed to wait for others to respond on asynchronous platforms or wait for scheduled synchronous meetings, resulting in slowed communication. P27 (35, M) goes more into detail about this problem: *"Before, I would be pulling my laptop on being able to just show me what's going on here. Now, after I actually show him, so I send him things and see exactly what is going on. And then I got to wait for a response so I don't get an immediate response anymore. So it's a little harder and it'll be like a glass as other person. I know it takes a little bit longer. So I think the collaboration has definitely changed. It's definitely a little bit slower in that manner."* Many participants felt the same way. P22 (23, F), a student, said that working with other students from different time zones further burdened communication efficiency.

The slow communication impacted work productivity as a result. P32 (33, F) is an attorney: *"So with the internet, you're kind of restricted in what you can communicate, how quickly you can communicate, you have to wait for a return email or response."* P27 (35, M) reflected on how working in-person allowed for more spontaneous communication, noting that things *"move[d] a lot faster and a lot of things would get done."* P29 (26, F), a geometry teacher, said: *"I believe the in-person collaborations were much more effective, and you were able to get more done rather than sending emails back and forth. It's better and more efficient when teachers meet to collaborate on upcoming events because it takes less time and more people can participate."* P29 pointed out the slow communication of work from home and the disadvantage of organizing events and mobilizing people online.

**4.1.2. Changes in the Working Environment** A few participants noted the difference in ambience between working at home and working at the office. P31 (26, F), a software engineer, said that the office provided a more 'official' work environment mood: *"I think the biggest difference is if I work at office, there are a lot of co-workers and work environment is more official cause you are specifically there. So you can feel the atmosphere around you like people sit around you and they talk about the work related business and transactions and some tasks."* P4 (23, M), an e-trader, also noticed the difference in ambience: *"A home is more of a homey vibe. I get to see my family the whole time. But at work it's more of a serious work environment."* P16 (22, M), a data analyst, felt that the office ambience helped him stay interested and attentive in his work.

**Distraction From Household** For many participants, their home environment and family members were major distractions while working from home. At least eleven participants pointed out their frustrations with these distractions. P42 (33, F) and P27 (35, M) both noted that these distractions made it more difficult to work from home. P11 (21, F), a graphic design intern, shared an example of how the family distractions impacted her work: *"One situation was where one day she, one of my managers, sent me assignment. I didn't see it till two hours after because I was working on something, and I was busy. Then I had to step away from my desk, which I probably wouldn't have done if I wasn't in the office, but that was home. My parents need something, so I stepped away and then it was like hour later and I was like, oops, I shouldn't have done that."* Even though P5 (N.A, N.A), a medical advisor, felt productive working from home, they still felt that being at home had more distractions compared to being at the office: *"I feel productive working at home. I'm good. I'm satisfied with productivity, but maybe from the office it's a little bit more productive at the office because you have less distractions and everyone else is working."*

On the other hand, some participants enjoyed working from home and felt it enhanced their productivity. Their home provided a quieter working environment compared to the office (P31, 26, F) and coworkers would not be a distraction (P46, 23, M). P16 (22, M), a data analyst, found that working from home supported his preference to work alone at his own pace. Two participants shared reasons as to why they preferred working in their own home. P44 (21, F), a student, said her workstation at home was better than the one at the office and liked how she had her own space, which helped her organize her thoughts. P46 (23, M) felt more relaxed and focused in the comfort of his own home: *"I just walk downstairs to the study where I have my workstation and do all my meetings and communicate by email or the desk phone that I brought from work. I'm pretty much more relaxed working at home just because I can be in my own environment, and I feel like I can focus better without any of the distractions from just like a normal workspace where people are just walking around and talking and getting other things done."*

**Lack of Facilities and Hands-on Experience** Work that cannot be done remotely at home had a harder time being able to continue during the pandemic due to the lack of necessary facilities and hands-on experience. Participants P15 (24, F), a production engineering intern, and P19 (21, M), a coding teacher, said that the lack of equipment and facilities made their work much more difficult. P20 (24, M), a reliability engineer, said that it was much harder to do his job because it requires traveling and meeting with clients. P7 (22, F), a process engineer, felt frustrated because a majority of her work took place outside, so working from home slowed her progress: *"I can't actually go to the plant to see if my projects are working."* In an educational setting, the lack of facilities impacted people's educational experiences as well. A high school science teacher, P17 (55, F), noted that doing hands-on activities were difficult to do without being at the lab. A university lecturer, P38 (29, M), said that Canvas, the platform he was using for classes, was very limited in teaching technical topics. Working remotely brought other small limitations, for example, signing paperwork has become more difficult and complicated (P46, 23, M).

Some missed having a more hands-on experience for their work. P46 talked about how the inability of being more hands-on was a disadvantage: *"So I'm a really visual person and hands on, so already going completely online kind of puts me at a disadvantage really. So when I'm explaining something to someone, it gets a little harder to communicate exactly what you need when you're doing virtually versus just writing or drawing it out."* Other participants also noticed the lack of having a 'complete' experience, like P15 (24, F), a production engineering intern who felt upset that she was unable to get the full internship experience she was expecting.

## 4.2. Adaptation of Work Practices

The second research question looks into the work practices participants adapted as a result of the changes in communication and environment. They adapted to the changes in five ways: expanding their technology portfolio, shifting their mental state, pushing themselves with hard deadlines, forcing improvements in communication and collaboration skills, and forcing more physical activities into their routine.

### 4.2.1. Expand Technology Portfolio

**Learn New Technology** Participants had to learn new technology, particularly new software, in order to be able to work remotely. For some, their organization facilitated the technology setup and users had only to learn to access it, but for others, they had to find the right technology in an ad-hoc manner. P12 (51, M) worked in a cybersecurity group for a bank: *"I think as a technology set up that we have at my organization, most of the pieces were already in place. So from remote access to, security log on, using multi-factor authentication, video conferencing systems as I've been in place...So I think the technology problem that the organization had made it easy."*

Seven participants expressed that they had to explore and learn new technologies in order to meet the work from home requirements. P27 (35, M) worked for a small startup and said he was pushed to use online software like JIRA to collaborate. P44 (21, F) was a teaching assistant for a dance class and shared her experience with technology setup: *"My dance teacher and I had to like figure out Zoom by ourselves, which was quite a feat. But we figured it out. I've definitely worked with her for like pretty long hours figuring everything out and making sure the students are satisfied and all that."*

**Diversify Communication Channels** Participants used the large array of technology available to make up for any lost communication methods (e.g. a virtual happy hour with coworkers on Google Hangouts - P40). For work, participants chose technology platforms based on what their fellow colleagues used. P44 (21, F) said, *"When I'm doing personal work with my group, if they all have iPhones it's usually FaceTime, and if they don't have iPhone, we usually go to Skype."*

Participants also chose different technologies based on the relationship they had with their colleagues. P15 (24, F) was a production engineering intern: *"The interns and I have like, a little group chat on iMessage so we usually use that if we have any questions between ourselves. We do have a GroupMe with our boss but we don't use it too often, usually when we don't get a good connection on Skype or something."* Having a separate interns-only group chat helped them speak more comfortably with each other without pressure from their boss. Similarly, P45 (21, F), a student, explained how she used social media to ask her classmates/friends for help, instead of using technologies that are designed for work collaboration: *"I use Zoom, WebEx, Google classroom with Google docs and email. We talk through email sometimes. If my classmate is one of my friends, then I would have had her or him on my social media. So I would text him, message him or anything if I need anything."*

*4.2.2.* **Shift in Mental State** Because of the inescapable nature of working at home during the pandemic, participants reported that they had to shift their mental state and accept the situation for what it was. P10 (46, M), a library employee, shared that the most difficult part of the transition was keeping a healthy mental state during the isolation: *"I can say that it will be easy but is difficult because of the mental part, being isolated in an environment for a long period of time and not having to interact with any human, and that's the difficult part. You know, anything in life when things change, it doesn't happen overnight... So it's difficult in a sense to adjust to it mentally and physically but psychologically."*

The shift was not easy and affected people's moods. P45 (21, F) is a social butterfly who expressed her struggles with accepting the isolation as the new norm: *"I was excited. I was expecting it to be easier, but after going through it for like the past four weeks, I feel like I can't do it anymore. It's getting really tough. It's like the social distancing part which is really tough. It is really hard for me because I'm not an introvert...It was going to be hard going back to normal life."* They had to learn how to adapt. P16 (22, M) said that working from home was initially boring, but he learned to adapt by being productive.

**Rely More on Independence** Four participants reported that they had to adjust their mindset to be more individualistic and independent due to the communication change. P11 (21, F), a graphic design intern, noted that working from home reduced her productivity and did not provide much support: *"Even though I try to be productive, I still feel like I can bounce ideas off of someone, talk to others and there's always like that support. But at home it's kind of like you figure it out on your own."* P4 (23, M) was an e-trader for a trading group who transitioned from working with a team to working alone: *"When I used to work at the office, it was more based on a team teamwork basis. So nowadays it's just me studying and making choices on my own."*

**Need New Mindset to Work with Clients** Three participants reported that they had to develop a new mindset for working with clients. For example, P26 (49, F) was an insurance agent who had to accommodate for older, non-tech-savvy customers to overcome challenges that would have been easier in-person: *"Since I deal with a lot of older people, it's harder for them to deal with technology and, for instance, if I need them to sign a paper, it's harder for them to scan and send it since they're not so tech-savvy."* Similarly, P36 (N.A, N.A) was a product analyst who usually resolved issues in-person, but now has to help clients remotely: *"So when we were working in person, we used to go to client site and help them resolve the issues... and now everything has become remotely."*

*4.2.3.* **Push Self with Hard Deadlines** Without the constant surveillance from their supervisors and colleagues, participants instead used hard deadlines as motivation to stay focused on their work. P1 (21, M) explained, *"I try to get everything done as best as I could, even though you tend to ease up a little bit more because you're in your own comfort, you're in your own house, but at the end of the day it's still your job and you have to do it. So I have to make sure that I get it done and have to get it done."* Sometimes, participants had to push themselves to maximize the use of their work time, even though their supervisor provided a lot of flexibility through loose deadlines and minimal updates. P7 (22, F), a process engineer, said, *"I guess I would just give myself hard deadlines. Like I have to get this done by Tuesday, I have to get this done by 2:00 PM. That's just my aspect."*

**Use Planning Technology** Eight participants specifically mentioned using productivity tools to help manage their work life. For example, P16 (22, M) used Google Calendar to track deadlines in a timely manner, and P18 (25, M) used a notebook to plan his day. P1 (21, M) was a receptionist at a dental office: *"I actually use a schedule, a calendar. I have whatever

*I have to do for my daily tasks, I would put in the calendar and then I'd just go based on what I have to do. I just go one by one and try to finish as fast as possible."* Similarly, P6 (23, M) was an assistant project manager who used calendar alarms to *"set up meetings"*.

#### 4.2.4. Force to Improve Communication and Collaboration Skills
The changes in communication also pushed participants to improve their personal skills due to the limitations of current collaboration methods.

**Improve Communication Skills** The limitations of communicative technology required participants to focus more on their communication skills (P16, 22, M) because meetings were more structured, making it easier to plan what to say. P42 (33, F) was a senior tax reporting and company analyst and said having online meetings allowed her to talk more freely without any interruptions. P12 (51, M) had to get used to having one person speak at a time, whereas in-person collaboration allowed for overlapping conversations: *"I mean on a video call, really only one person can speak at a time so that change and going from in-person, physical to a video conferencing, we've had to adjust to one person speaking at a time. And you know, you cannot talk at the same time."*

**Learn to Collaborate More Effectively** Because online meetings were more structured, several participants had to learn how to collaborate more effectively, to maximize on the limited time they had with their coworkers. P44 (21, F) used Zoom to teach an online dance class: *"So we would have one student perform and then like we'd pause the performance and like we talk about the performance. It was a lot more collaborative than it would have been in person."*

#### 4.2.5. Force to Have More Physical Activities
Sitting in front of a desktop at home for long periods of time negatively impacted people's emotions and physical health. Some consequences of inactivity included straining eyes or moving around less (P17, 55, F). As a result, participants forced themselves to do more exercises, like walking around the house or in the yard. P41 (36, F), a manager in risk adjustment for a healthcare company, depended on others to push her to do workout at home:*"At the gym I use a lot of different types of equipment. Whereas at home I can only do self workouts. It was awkward, and I wasn't able to zone out running, for example. My brother kept checking in on me."*

**Frequent and Flexible Breaks** The flexibility and lack of surveillance allowed participants to take more breaks. Though the breaks were *"mundane"* (P16, 22, M), participants frequently took them to relax (e.g., taking naps-P15) or to cope with stress. P1 (21, M) as a receptionist said, *" Because I feel like the more stressed I am, the more likely I am to make a mistake and want to avoid that. So every hour I'd take a five minute break, maybe go eat something, use the bathroom if I need to wash my face, any of that."*

## 5. Discussion

Telecommuters have to compromise on the limited technological infrastructure and passively adapt to them. The passive transition is not the expected performance or a long-term solution. How we can overcome the shortcomings of technology infrastructure in the home environment and motivate people to actively communicate should be further investigated. Prior work explains the transition process by discussing the shift of technological affordances and environmental affordances [3]. Our findings supplement their work with specific changes. For example, they found that COVID-19 has shifted affordances dramatically, requiring a new pattern of communication in terms of the frequency, length, and the style. We found that the asynchronous communication requires more writing than verbal and increases the communication volume and decrease the productivity. The communication change also further demotivates the initiative to interact with others to some extent because information and communication can be saved and archived. In this sense, it decreases the communication frequency, which is also supported by the mental shift of relying more on working independently.

The adaptation strategies involve physical, psychological, and practical activities with limited technological intervention. Participants only used planning technology to push themselves for task management. The difficulties and limitations of current work and collaboration technology pushed telecommuters to improve their work practices. For example, people gained more personal skills related to technology use and team collaboration, a rare but positive impact of being forced to work from home, which contrasts with most literature that focuses on the negative impacts [6, 21, 25].

The adaptation strategies can help people overcome communication challenges, but not the challenges caused by the changes in environment. Telecommuters express concerns about household distractions and lack of facilities and hands-on experiences in the home environment, with no specific strategy to help cope with these problems. Enforced work from home telecommuters have to accept the situation and delay their work with the risk of being infected if they visit the office. People may think it is a temporal situation, but what if the pandemic continues and prolongs for several years? Understanding how to cope with the limitations of the home environment and knowing what technological intervention can mitigate its impact, are critical for operation in relevant industries.

### 5.1. Limitations and Future Work

For this study, we only chose people who can work from home during the pandemic, not including those who still need to work in the office or the field, for example, nurses and doctors in hospitals [29]. Furthermore, this study focuses more on the practical changes and adaptations of working at home, not the psychological impacts. We did highlight the psychological strategies people used to cope with the changes; however, we did not explore this topic further during interviews to learn the cognitive process of how they shifted their mental state, requiring further investigation. Some noticeable differences should be highlighted and could be interesting topics for future research. For example, is there a difference in performance between those with prior remote working experience and those with no experience? To what extent? How do we better support those who have no experience? Additionally, some organizations provided good work infrastructure and resources for their employees. How does the infrastructural difference influence telecommuters' performance? Last, The data collection completed in the early stage of the pandemic. It is an unusual situation for the participants to use technology for teleworking. They might experience difficulty in using technology. Nowadays, they might adapt to technology use and changed their behaviors and perceptions towards the technology. Future work can also follow this study to explore the dynamics of perception of technology use during the pandemic.

### 6. Conclusion

As a result of the COVID-19 quarantine and social distancing mandates, people underwent many practical changes in communication and environment to be able to work from home. Most were able to adapt and accommodate their work practices to support this new normal lifestyle. However, these adaptations were not a sustainable solution for the long run. There are many limitations that hindered people's ability to work at home at the same capacity as in the office. To make work from home a more sustainable practice for the future, we would need to solve the current limitations of remote work technology or develop a work structure that allows for more flexibility like a hybrid in-person/remote work schedule.

## A. Appendix

Table 1. Demographic and Industry of Participants

| Participants | Occupation & Industry | Age | Race | Gender |
|---|---|---|---|---|
| P1 | Receptionist at a Dental Office | 21 | White | Male |
| P2 | Customer Services Representative | 22 | White | Female |
| P3 | Scheduling Manager | 19 | Asian | Male |
| P4 | E-trader | 23 | White | Male |
| P5 | Medical Advisor | N/A | N/A | N/A |
| P6 | Assistant Project Manager | 23 | White | Male |
| P7 | Process Engineer | 22 | Asian | Female |
| P8 | Emergency Medical Technician | 19 | Asian | Male |
| P9 | Desk Consultant and Fitness Employee | 31 | Hispanic | Male |
| P10 | Library Staff | 46 | African American | Male |
| P11 | Graphic Design Intern | 21 | Asian | Female |
| P12 | Cybersecurity Technology At a Bank | 51 | Asian | Male |
| P13 | Pediatric Newborn Research Assistant | 23 | Asian | Female |
| P14 | Programmer & Business Owner | 29 | Asian | Male |
| P15 | Production Engineering Intern | 24 | Asian | Female |
| P16 | Data Analyst | 22 | Asian | Male |
| P17 | High School Science Teacher | 55 | Asian | Female |
| P18 | Software Developer | 25 | White | Male |
| P19 | Coding Teacher | 21 | Asian | Male |
| P20 | Reliability Engineer | 24 | Asian | Male |
| P21 | Technology Co-op | 22 | Asian | Female |
| P22 | Student | 23 | Asian | Female |
| P23 | Unemployed | 29 | White | Male |
| P24 | Senior Research Analyst in Healthcare | 26 | African American | Male |
| P25 | RF Engineer for Corporate | 27 | Asian | Male |
| P26 | Insurance Agent | 49 | Hispanic | Female |
| P27 | Small Startup | 35 | White | Male |
| P28 | Talent Coordinator | 23 | African American | Female |
| P29 | Geometry Teacher | 26 | African American | Female |
| P30 | Software Engineer | 31 | Asian | Male |
| P31 | Software Engineer | 26 | Asian | Female |
| P32 | Attorney | 33 | White | Female |
| P33 | Software Engineer | 26 | Asian | Male |
| P34 | Corporate Office Manager for Car Wash | 30 | White | Female |
| P35 | Student | N/A | N/A | N/A |
| P36 | Product Analyst | N/A | N/A | N/A |
| P37 | Student | 22 | Asian | Female |
| P38 | University Lecturer | 29 | White | Male |
| P39 | Ph.D. student | 25 | Asian | Male |
| P40 | Product Owner in Customer Technology Team | N/A | N/A | N/A |
| P41 | Manager in Risk Adjustment for Healthcare Company | 36 | African American | Female |
| P42 | Senior Tax Reporting and Company Analyst | 33 | Hispanic | Female |
| P43 | Masters Student | N/A | N/A | N/A |
| P44 | Student & Dance Teacher | 21 | Asian | Female |
| P45 | Student & Substitute | 21 | Asian | Female |
| P46 | Network Administrator, Developer, Recruiter | 23 | Asian | Male |